\documentclass{article}

\usepackage{arxiv}

\usepackage[utf8]{inputenc}
\usepackage[T1]{fontenc}
\usepackage{amsmath,amssymb}
\usepackage{graphicx}
\usepackage{subcaption}
\usepackage{booktabs}
\usepackage{hyperref}
\usepackage{url}
\usepackage{array}
\usepackage{multirow}
\usepackage{xcolor}
\usepackage{algorithm}
\usepackage{algpseudocode}
\usepackage[numbers]{natbib}
\usepackage{doi}
\usepackage{threeparttable}
\usepackage{pifont}
\usepackage{microtype}
\usepackage{float}
\usepackage{placeins}
\newcommand{\cmark}{\ding{51}}
\newcommand{\xmark}{\ding{55}}

\title{T5Gemma-TTS Technical Report}

\author{
  Chihiro Arata\textsuperscript{1},
  Kiyoshi Kurihara\textsuperscript{2,3} \\[6pt]
  \small\textsuperscript{1}Third Intelligence, Inc. \\
  \small\textsuperscript{2}Matsuo Institute, Inc. \\
  \small\textsuperscript{3}Department of Technology Management for Innovation, \\
  \small Graduate School of Engineering, The University of Tokyo \\[4pt]
  \small\texttt{chihiro.arata@third-intelligence.com},
  \texttt{kiyoshi.kurihara@weblab.t.u-tokyo.ac.jp}
}

\date{\today}

\begin{document}
\maketitle

\begin{abstract}
Autoregressive neural codec language models have demonstrated remarkable
zero-shot voice cloning capability, yet dominant decoder-only architectures
treat input text as a prefix that competes with the growing audio sequence
for positional capacity, causing text conditioning to dilute over long
utterances.
We present \textbf{T5Gemma-TTS}, an encoder-decoder codec language model
that routes bidirectional text representations through cross-attention at
every decoder layer, maintaining persistent and structured text conditioning
regardless of output length.
Built on the T5Gemma pretrained encoder-decoder backbone
(2\,B encoder\,+\,2\,B decoder, 4\,B parameters in total),
our model inherits rich linguistic knowledge without requiring phoneme
conversion, enabling direct subword-level text processing.
To overcome the inherent difficulty of duration control in autoregressive
generation, we integrate \textbf{Progress-Monitoring Rotary Position
Embedding (PM-RoPE)} into all 26 cross-attention layers, injecting
normalized generation-progress signals that allow the decoder to continuously
track its position relative to the target speech length.
Trained on approximately 170{,}000 hours of multilingual speech in English,
Chinese, and Japanese, T5Gemma-TTS achieves a statistically significant
speaker similarity (SIM) advantage
on Japanese (SIM\,=\,0.677 vs.\ XTTS\,v2~0.622; non-overlapping 95\%
confidence interval (CI))
and the numerically highest Korean speaker similarity (SIM\,=\,0.747)
despite Korean being absent from training---though this Korean advantage
over XTTS\,v2 (0.741) is not statistically conclusive (overlapping CI).
T5Gemma-TTS also records the numerically lowest Japanese character error
rate (CER) of 0.126
among five competitive baselines, though the ranking should be interpreted
cautiously due to partial CI overlap with Kokoro.
English results on LibriSpeech should be treated as an upper-bound estimate,
as LibriHeavy (one English training source) is a superset of LibriSpeech.
A configuration analysis comparing PM-RoPE-enabled and PM-RoPE-disabled inference
on the same trained checkpoint shows that disabling PM-RoPE causes near-complete
synthesis failure (CER degrades from 0.129 to 0.982; duration accuracy (DA) drops
from 79\% to 46\%), demonstrating that PM-RoPE is essential for coherent
text-conditioned generation; methodological details and
caveats are discussed in Section~\ref{sec:ablation}.
Code and model weights are available at
\url{https://github.com/Aratako/T5Gemma-TTS}.
\end{abstract}

\keywords{text-to-speech, zero-shot voice cloning, encoder-decoder, PM-RoPE, multilingual evaluation}

\section{Introduction}
\label{sec:intro}

\paragraph{Zero-shot TTS and neural codec language models.}
Zero-shot text-to-speech (TTS) synthesis---the ability to clone an arbitrary
speaker's voice from a short reference clip without target-speaker
training---has advanced rapidly through the neural codec language model
(NCLM) paradigm~\cite{vallex2023,vallex22024,seedtts2024}.
By discretizing continuous speech waveforms into compact token sequences
via neural audio codecs~\cite{encodec2022,llasa2025}, NCLMs recast TTS as a
conditional language modeling problem, enabling high-fidelity voice cloning
at scale.
State-of-the-art systems such as VALL-E 2~\cite{vallex22024} and
Seed-TTS~\cite{seedtts2024} have achieved near-human naturalness on English
benchmarks, while non-autoregressive flow-matching models such as
F5-TTS~\cite{f5tts2024} and E2 TTS~\cite{e2tts2024} offer competitive
quality with efficient parallel decoding.

\paragraph{Limitations of decoder-only text conditioning.}
Despite this progress, the dominant decoder-only NCLM architecture harbors
a structural limitation in how it conditions on input text.
Decoder-only models concatenate the text token sequence as a prefix
to the audio token sequence and apply causal self-attention uniformly over
the combined sequence.
As generation proceeds, the audio sequence length $S$ grows while the text
prefix length $T$ remains fixed; consequently, text tokens occupy a
diminishing fraction of each attention window, and their influence on the
decoder's hidden states weakens as $S \gg T$.
This dilution effect becomes increasingly pronounced for long utterances---a
regime that is common in document-level TTS and audio book synthesis.
Encoder-decoder models circumvent this problem by design: the encoder
processes the full text bidirectionally and produces a fixed-size
representation matrix $\mathbf{H}^{\mathrm{enc}} \in \mathbb{R}^{T \times d}$,
which is routed to the decoder via dedicated cross-attention at
every layer.
Text conditioning therefore operates through a separate, persistent pathway
that is decoupled from the causal audio self-attention, preserving
full text context regardless of output length.

\paragraph{Duration control in autoregressive generation.}
Accurate duration control is an important practical consideration in
autoregressive speech generation.
Without explicit positional grounding, an autoregressive model has limited
signals for how far through the target speech it has progressed, which can
make matching a desired duration less consistent.
\citet{voicestar2025} explored this in an encoder-decoder setting via
Progress-Monitoring Rotary Position Embedding (PM-RoPE), which encodes
normalized generation progress into cross-attention queries and keys,
effectively informing the decoder of its current position relative to the
full target length.
While promising, VoiceStar~\cite{voicestar2025} was trained from scratch,
so it remains useful to examine whether the approach also scales and
generalizes across languages when initialized from a large pretrained
language model.

\paragraph{T5Gemma-TTS.}
We address both limitations by proposing \textbf{T5Gemma-TTS}, an
autoregressive encoder-decoder codec language model for multilingual
zero-shot TTS.
Our contributions are:

\begin{itemize}
  \item \textbf{Pretrained encoder-decoder foundation.}
    T5Gemma-TTS is built on T5Gemma~\cite{t5gemma2024}, a 2B$+$2B
    parameter encoder-decoder large language model (LLM) pretrained on
    large-scale multilingual text corpora via the Unifying Language Learning (UL2) objective.
    Initializing from this checkpoint transfers broad linguistic knowledge
    to the audio domain, obviates the need for phoneme conversion, and
    provides a strong multilingual subword vocabulary covering English,
    Chinese, and Japanese out of the box.

  \item \textbf{Integration and multilingual validation of PM-RoPE.}
    We integrate PM-RoPE~\cite{voicestar2025}---a duration-control mechanism
    originally proposed by VoiceStar for English-only TTS trained from
    scratch---into all 26 cross-attention layers of the T5Gemma decoder
    without modification to the mechanism itself.
    Our contribution is the first demonstration that PM-RoPE generalizes
    beyond English monolingual training: applied to a 4\,B-parameter
    pretrained multilingual backbone trained on $\sim$170k hours across
    three typologically diverse languages, PM-RoPE preserves its
    duration-control properties while substantially expanding language coverage.

  \item \textbf{Large-scale multilingual training across three typologically diverse languages.}
    T5Gemma-TTS is trained on approximately 170{,}000 hours of speech
    spanning English ($\sim$100k h; stress-timed, Indo-European),
    Mandarin Chinese ($\sim$50k h; tonal, Sino-Tibetan),
    and Japanese ($\sim$20k h; mora-timed, Japonic).
    The training corpus is constructed from two complementary resources:
    Emilia~\cite{emilia2024}, which provides multilingual open-source speech,
    and LibriHeavy~\cite{libriheavy2024}, which provides large-scale
    English read speech; together they form the 170k-hour training set.
    While this is narrower in scope than systems trained on ten or more
    languages, the typological diversity of the three training languages
    allows us to study cross-lingual generalization, as evidenced by
    competitive Korean performance (a language entirely absent from training).
    The term multilingual throughout this paper refers specifically
    to this three-language training scope.

\end{itemize}

Experimental evaluation on six languages
(Japanese, Mandarin Chinese, English,
Korean, French, and German)
shows that T5Gemma-TTS achieves a statistically significant speaker
similarity (SIM) advantage on Japanese
($\mathrm{SIM} = 0.677 \pm 0.016$ vs.\
XTTS\,v2 $0.622 \pm 0.016$; 95\% confidence interval (CI) non-overlapping)
and the numerically highest
Korean SIM ($0.747 \pm 0.029$; overlapping with XTTS\,v2
$0.741 \pm 0.010$) among all five evaluated systems
(F5-TTS~\cite{f5tts2024}, XTTS\,v2~\cite{casanova2024xtts},
CosyVoice\,2~\cite{du2024cosyvoice2}, Kokoro~\cite{kokoro2025}).
T5Gemma-TTS achieves the second-highest Chinese speaker similarity among
zero-shot voice cloning systems ($\mathrm{SIM} = 0.722 \pm 0.017$); only
F5-TTS records a higher SIM (0.864) but at the cost of severely degraded
intelligibility (character error rate (CER) $= 0.155$ despite Chinese being
a training language).
T5Gemma-TTS also records the numerically lowest Japanese CER (0.126),
though confidence intervals partially overlap with Kokoro.
English results on LibriSpeech should be read as an upper-bound
estimate, as LibriHeavy is a LibriSpeech superset.
T5Gemma-TTS substantially outperforms duration-uncontrolled baselines on
Duration Accuracy (0.79 vs.\ 0.46 without PM-RoPE).
The extended evaluation further reveals that T5Gemma-TTS achieves
competitive Korean CER (0.082) and SIM (0.747) despite Korean being
entirely absent from its training data, demonstrating the value of the
pretrained multilingual T5Gemma backbone for cross-lingual generalization.


\section{Related Work}
\label{sec:related}

\subsection{Neural Codec Language Models for TTS}

The emergence of discrete audio codecs as an intermediate representation has
enabled the reformulation of text-to-speech synthesis as a language modeling
problem.
VALL-E~\cite{vallex2023} pioneered this paradigm by conditioning a
decoder-only codec language model on EnCodec~\cite{encodec2022} tokens, achieving zero-shot voice
cloning from only three seconds of reference audio.
Subsequent work addressed robustness and quality:
VALL-E~2~\cite{vallex22024} introduced Repetition-Aware Sampling
(RAS)~\cite{vallex22024} and
Grouped Code Modeling (GCM)~\cite{vallex22024}, reaching human parity on
LibriSpeech~\cite{panayotov2015librispeech} and VCTK benchmarks.
Seed-TTS~\cite{seedtts2024} further improved expressiveness through
speech-level self-distillation and reinforcement learning from human feedback,
while also proposing a non-autoregressive diffusion-based variant.

\paragraph{Llasa and XCodec2.}
Llasa introduced XCodec2~\cite{llasa2025}, a single-codebook neural audio
codec operating at 50~Hz with vocabulary size 65{,}536, and applied it to
a decoder-only LLaMA~\cite{touvron2023llama}-based codec language model trained with scaled
compute on English speech.
T5Gemma-TTS adopts XCodec2 as its audio tokenizer, leveraging its
single-codebook design to keep the audio token sequence short.
However, T5Gemma-TTS diverges from Llasa architecturally by replacing the
decoder-only backbone with a pretrained encoder-decoder model (T5Gemma),
enabling bidirectional text encoding via cross-attention at every decoder
layer and supporting multilingual subword tokenization without phoneme
conversion.

All of these systems adopt a decoder-only backbone in which the input
text is prepended as a prefix to the audio token sequence.
While effective, this approach reduces the textual signal to a fixed-length
prefix that can be effectively ``forgotten'' as the audio sequence grows longer.
T5Gemma-TTS takes a different stance: by adopting an encoder-decoder
architecture, the encoder output is injected into every decoder layer
via cross-attention, maintaining strong and structured text conditioning
throughout the full generation trajectory.

\subsection{Encoder-Decoder Architectures in TTS}

Encoder-decoder sequence-to-sequence models have been a mainstay of TTS
for nearly a decade, starting with
Tacotron-style systems~\cite{shen2018tacotron2} that map phoneme sequences to
mel-spectrograms through attention.
FastSpeech~2~\cite{fastspeech22021} refined this paradigm by explicitly
predicting duration, pitch, and energy through a variance adaptor, eliminating
the instability of autoregressive attention-based alignment and enabling
fine-grained prosody control.
However, these classical encoder-decoder systems rely on phoneme-level inputs
and separately trained duration predictors, adding pipeline complexity.

More recently, VoiceStar~\cite{voicestar2025} proposed applying the
encoder-decoder codec language model framework to zero-shot TTS and introduced
Progress-Monitoring Rotary Position Embedding (PM-RoPE)~\cite{voicestar2025} to inject normalized
generation progress into cross-attention.
PM-RoPE allows the model to continuously track how far through the target
audio it has progressed, enabling reliable duration control and extrapolation
to speech lengths beyond the training distribution.
VoiceStar was trained from scratch on an English-only corpus.

T5Gemma-TTS directly adopts PM-RoPE from VoiceStar~\cite{voicestar2025}
(Section~\ref{sec:pm_rope}) with no modification to the mechanism itself.
A notable difference from VoiceStar is the text representation:
VoiceStar converts input text to phoneme sequences, which naturally
provides a monotonic alignment between text and audio positions
well-suited for PM-RoPE's progress-monitoring mechanism.
T5Gemma-TTS instead feeds subword tokens directly from the T5Gemma
SentencePiece~\cite{kudo2018sentencepiece} tokenizer, sacrificing the
monotonic phoneme--audio correspondence in favor of two practical benefits:
(1) avoiding the engineering cost of language-specific phonemizers for
each target language, and (2) preserving the pretrained embedding weights
from the T5Gemma backbone, which encode rich multilingual semantic
information.
The effect of this phoneme-vs-subword choice on PM-RoPE's
duration control effectiveness has not been ablated in this work and
remains an open question for future investigation.
Our contribution with respect to PM-RoPE is therefore not the invention of
the technique but rather its application to a large pretrained
encoder-decoder backbone (T5Gemma~\cite{t5gemma2024}) with subword
(rather than phoneme) input, and its empirical
validation at substantially larger data scale ($\sim$170k hours) and across
multiple typologically diverse languages (English, Chinese, Japanese).
This large-scale multilingual training experiment provides the first evidence that PM-RoPE generalizes
beyond the English monolingual regime in which it was originally proposed.

\subsection{Flow-Matching and Diffusion TTS}

As an alternative to autoregressive codec language models~\cite{vallex2023,vallex22024,seedtts2024},
flow-matching~\cite{f5tts2024,e2tts2024} and
diffusion-based systems operate in continuous acoustic space.
E2~TTS~\cite{e2tts2024} demonstrated that a minimal flow-matching model
trained on the audio-infilling task, without any duration predictor or
phoneme aligner, could achieve competitive zero-shot TTS.
F5-TTS~\cite{f5tts2024} refined this approach with a Diffusion Transformer
(DiT)~\cite{peebles2023dit} backbone and ConvNeXt~\cite{liu2022convnext}-based text representation, achieving
state-of-the-art naturalness on English benchmarks.
CosyVoice~2~\cite{du2024cosyvoice2} adopted a hybrid strategy combining a
large-scale language model for semantic token prediction with a
flow-matching-based acoustic model for high-quality multilingual synthesis.

While these non-autoregressive models excel in naturalness and inference
speed, they typically offer limited explicit duration control; duration is
governed by an implicit coupling between text padding length and mel-spectrogram
length rather than by an explicit, user-specified target duration.
T5Gemma-TTS addresses this gap by providing controllable and predictable
output durations through PM-RoPE, while retaining the naturalness benefits of
large-scale pretraining.

\subsection{Duration Control in TTS}

Explicit duration control has been studied extensively in the non-autoregressive
TTS literature.
FastSpeech~2~\cite{fastspeech22021} predicts frame-level phoneme durations
from a duration predictor trained with Montreal Forced Aligner~\cite{mcauliffe2017montreal} annotations.
MaskGCT~\cite{maskgct2024} enables coarse-grained total duration control
by conditioning the masked generative model on the target total token count.
In the autoregressive setting, VoiceStar~\cite{voicestar2025} showed that
PM-RoPE provides fine-grained, continuous duration signals that generalize to
long-form (20--50 seconds) synthesis without requiring explicit duration labels
during training.
T5Gemma-TTS adopts the PM-RoPE mechanism proposed by
VoiceStar~\cite{voicestar2025} verbatim and pairs it with a
phoneme-count-based duration estimator (Section~\ref{sec:duration}) to
determine the target token length at inference time, enabling both automatic
and user-specified duration control.
The novel contribution of T5Gemma-TTS is the combination of PM-RoPE with a
large multilingual pretrained encoder-decoder backbone, which we show
preserves its duration-control properties while substantially expanding
language coverage.

\subsection{Multilingual TTS}

Multilingual zero-shot TTS is increasingly important for global deployment.
CosyVoice~2~\cite{du2024cosyvoice2} achieves multilingual synthesis across
multiple languages using supervised semantic tokens and flow-matching acoustics,
while Seed-TTS~\cite{seedtts2024} demonstrates multilingual voice cloning at
scale.
Recent work has also explored generalization to languages entirely absent
from training: \citet{saeki2024extending100} extended speech synthesis to
over 100 languages---including 50 with no transcribed training data---by
combining multilingual speech--text pretraining with minimal adaptation,
achieving intelligible zero-shot synthesis in the majority of unseen languages.
T5Gemma-TTS leverages the T5Gemma backbone~\cite{t5gemma2024}---an
encoder-decoder adaptation of Gemma~2 with broad multilingual text
pretraining---and extends this to audio generation by training on
$\sim$170k hours of speech spanning three typologically diverse languages:
English (Indo-European, stress-timed), Mandarin Chinese (Sino-Tibetan, tonal),
and Japanese (Japonic, mora-timed), drawn from the
Emilia~\cite{emilia2024} and LibriHeavy~\cite{libriheavy2024} datasets.
While this coverage is narrower than systems explicitly trained on ten or more
languages (e.g., XTTS\,v2 supports 16 languages), the focus on three
phonologically diverse training languages allows us to study cross-lingual
generalization---as demonstrated by competitive Korean performance despite
Korean being absent from training.
The use of a subword tokenizer (rather than a language-specific phoneme
converter) facilitates straightforward extension to additional languages
in future work.

\section{Method}
\label{sec:method}

\subsection{Model Architecture}
\label{sec:arch}

\begin{figure}[!t]
  \centering
  \includegraphics[width=\linewidth]{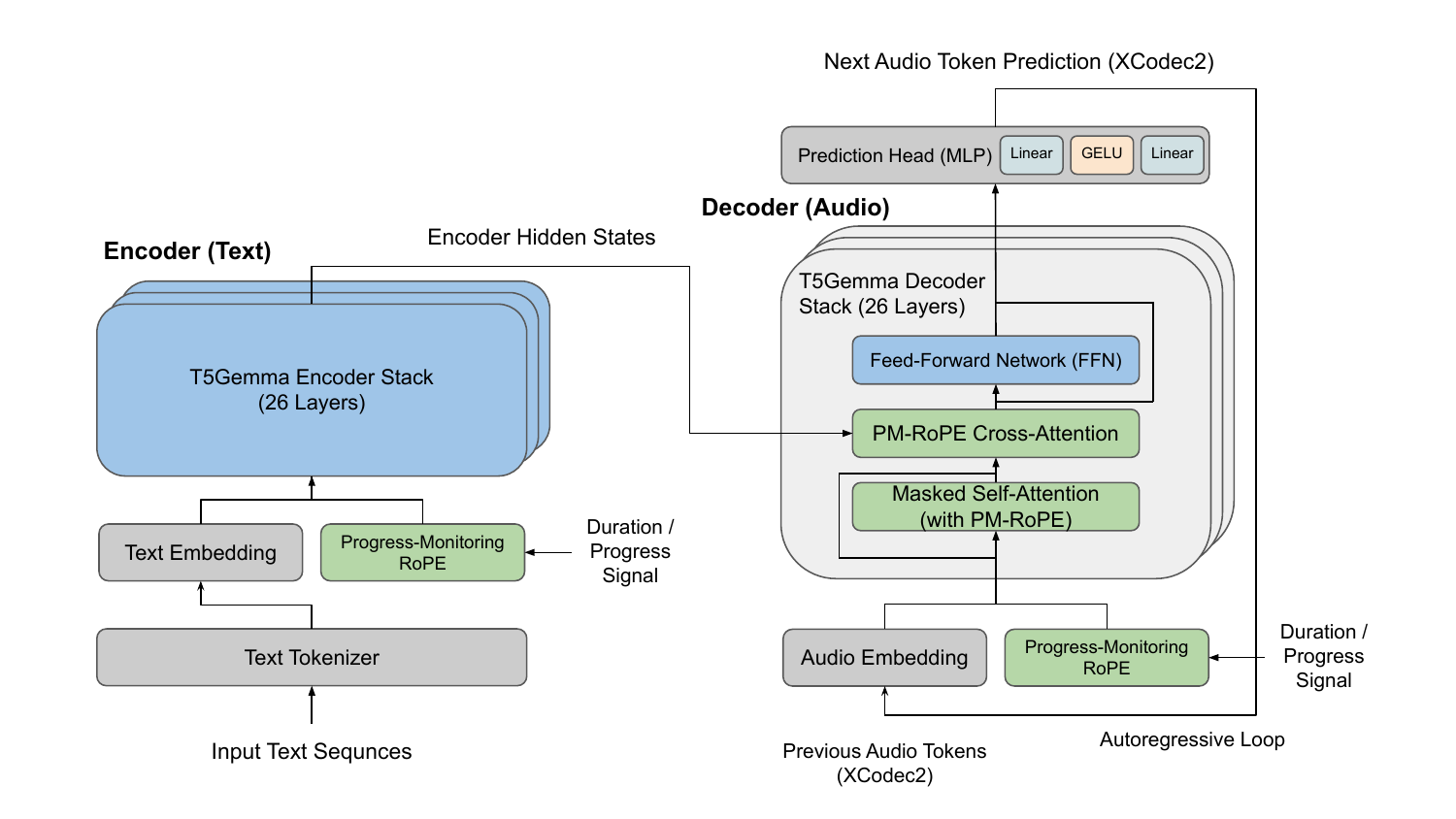}
  \caption{Overall architecture of T5Gemma-TTS.
  The T5Gemma encoder processes input text bidirectionally and produces
  contextualized representations, which are injected into every decoder
  layer via PM-RoPE cross-attention.
  The decoder autoregressively generates XCodec2 audio tokens conditioned
  on both the encoder output and a reference speech prompt.}
  \label{fig:architecture}
\end{figure}

Figure~\ref{fig:architecture} illustrates the overall architecture of
T5Gemma-TTS.
The model is an autoregressive sequence-to-sequence model that maps a text
sequence $\mathbf{x} = (x_1, \dots, x_T)$ and an optional reference speech
waveform $\mathbf{r}$ to discrete audio tokens
$\mathbf{y} = (y_1, \dots, y_S)$.

Audio tokens are produced by XCodec2, the neural audio codec
released as part of the Llasa project~\cite{llasa2025}.
The T5Gemma-TTS codebase uses a Japanese fine-tuned variant of XCodec2
as its default codec; the encoder is identical to the original
\texttt{HKUSTAudio/xcodec2}~\cite{llasa2025} (ensuring tokenization compatibility),
while the decoder has been fine-tuned on Japanese speech.
This may slightly disadvantage non-Japanese reconstruction quality
due to the decoder's language-specific tuning; however, since our
evaluation metrics (CER/WER, SIM) operate on the generated
waveform (decoded by the fine-tuned decoder), this effect is consistent
across all evaluated languages.
The codec operates at 50~Hz with a single codebook of vocabulary size $V = 65{,}536$.
We select XCodec2 over EnCodec~\cite{encodec2022} and DAC (Descript Audio Codec)~\cite{kumar2023dac} for three reasons.
First, its single-codebook design avoids the interleaved multi-codebook
prediction required by EnCodec's residual vector quantization (RVQ)~\cite{zeghidour2021soundstream} layers, keeping the decoder's generation
task equivalent to standard next-token language modeling.
Second, its 50~Hz frame rate provides a favorable trade-off between temporal
resolution and sequence length---lower than DAC's 86~Hz, reducing the audio
token sequence length for a given utterance duration and thereby lowering
memory and compute requirements.
Third, its vocabulary size of 65{,}536 offers substantially finer-grained
acoustic quantization than EnCodec's 1{,}024-code single-codebook variant,
enabling higher reconstruction fidelity without introducing residual codebooks.

\paragraph{Text encoder.}
The T5Gemma encoder \cite{t5gemma2024} processes input text through 26 transformer
layers with model dimension $d = 2{,}304$, producing contextualized representations
$\mathbf{H}^{\mathrm{enc}} \in \mathbb{R}^{T \times d}$.
The encoder is initialized from the UL2-pretrained T5Gemma-2b-2b checkpoint,
enabling direct use of subword tokenization without phoneme conversion.

We select T5Gemma~\cite{t5gemma2024} over alternative multilingual encoders
(e.g., mT5, NLLB-200) for three reasons:
(1) the 2\,B parameter scale provides richer representations than mT5-Large
($\approx$580\,M parameters);
(2) the UL2 pretraining objective combines denoising, prefix-LM, and causal-LM
losses, yielding bidirectional representations well-suited for text encoding
in a sequence-to-sequence TTS model;
and (3) the SentencePiece~\cite{kudo2018sentencepiece} vocabulary natively covers Japanese and Chinese
subwords without language-specific phoneme conversion, simplifying the
multilingual input pipeline.

\paragraph{Audio decoder.}
The decoder mirrors the encoder depth (26 layers, $d = 2{,}304$) and
autoregressively predicts the next audio token from:
(1) an audio embedding $\mathbf{E} \in \mathbb{R}^{V' \times d}$
(where $V' = V + 5$, with the five special tokens being
\texttt{<bos>}, \texttt{<eos>}, \texttt{<pad>}, a silence token, and a
prompt-separator token used to delimit the reference-audio prefix from
the generation target), and
(2) the encoder output $\mathbf{H}^{\mathrm{enc}}$ via cross-attention.
A linear projection layer with Gaussian Error Linear Unit (GELU) activation maps decoder hidden states to
token logits.
Input audio is resampled to 16\,kHz before XCodec2 tokenization;
the reference prompt waveform is prepended to the generation target in the
token sequence, separated by the prompt-separator token.

\subsection{Progress-Monitoring RoPE Cross-Attention}
\label{sec:pm_rope}

In standard cross-attention, the decoder query and encoder key vectors carry
no information about their respective positions within their sequences.
This makes it difficult for the model to track how far through the target
speech it has progressed, hampering duration control.

We apply \textbf{PM-RoPE}~\cite{voicestar2025} to all 26 cross-attention layers of the
T5Gemma decoder.
The mechanism is adopted from VoiceStar without modification;
our contribution is its integration into a large multilingual pretrained
backbone and its empirical validation across multiple languages.
PM-RoPE addresses the duration control problem by embedding relative
generation progress into the cross-attention mechanism.
For decoder position $j \in \{0, \dots, S-1\}$ and encoder position
$i \in \{0, \dots, T-1\}$, we define progress position IDs as:

\begin{equation}
  p^{\mathrm{dec}}_j = \frac{j}{S - 1} \cdot s, \qquad
  p^{\mathrm{enc}}_i = \frac{i}{T - 1} \cdot s,
  \label{eq:progress_ids}
\end{equation}

where $s = 2{,}000$ is a fixed scaling constant.
Two independent RoPE modules---one for decoder queries and one for encoder
keys---apply rotary position embeddings using these progress IDs:

\begin{align}
  \tilde{\mathbf{q}}_j &= \mathrm{RoPE}_{\mathrm{dec}}\!\left(
    \mathbf{W}_Q \mathbf{h}^{\mathrm{dec}}_j,\; p^{\mathrm{dec}}_j
  \right), \label{eq:pm_q} \\
  \tilde{\mathbf{k}}_i &= \mathrm{RoPE}_{\mathrm{enc}}\!\left(
    \mathbf{W}_K \mathbf{h}^{\mathrm{enc}}_i,\; p^{\mathrm{enc}}_i
  \right). \label{eq:pm_k}
\end{align}

The resulting attention scores between $\tilde{\mathbf{q}}_j$ and
$\tilde{\mathbf{k}}_i$ encode the alignment between the decoder's current
progress and each encoder position, enabling the model to attend to text
tokens proportional to its generation stage.
PM-RoPE cross-attention replaces the standard cross-attention in all 26
decoder layers.

\paragraph{Inference-time duration control.}
At inference, the target speech duration $\hat{D}$ (in seconds) is estimated
from the reference audio and target text (Section~\ref{sec:duration}).
The total target token count is $\hat{S} = \lfloor \hat{D} \times 50 \rfloor$.
At each decoding step $j$, the decoder progress ID is set to
$p^{\mathrm{dec}}_j = j / (\hat{S} - 1) \cdot s$, providing dynamic,
length-conditioned positional signals.

\subsection{Duration Estimation}
\label{sec:duration}

Given target text $\mathbf{x}$ and reference speech $\mathbf{r}$ with
transcript $\mathbf{r}_{\text{txt}}$, the target duration is estimated as:

\begin{equation}
  \hat{D} = \frac{D_{\mathrm{ref}}}{N_{\mathrm{ref}}} \cdot N_{\mathrm{tgt}},
\end{equation}

where $D_{\mathrm{ref}}$ is the reference speech duration, $N_{\mathrm{ref}}$
and $N_{\mathrm{tgt}}$ are the phoneme counts of the reference and target
texts, respectively.
Phoneme counts are obtained via \texttt{espeak-ng} for English,
\texttt{pyopenjtalk}~\cite{yamamoto2018pyopenjtalk} for Japanese, and Unicode character count for Chinese
(one character $\approx$ one mora/syllable).
When no reference is available, language-specific default rates are used:
$\Delta_{\mathrm{EN}} = 0.085$~s/phoneme,
$\Delta_{\mathrm{JA}} = 0.10$~s/phoneme,
$\Delta_{\mathrm{ZH}} = 0.27$~s/character.
For out-of-training languages (Korean, French, German) evaluated in
Section~\ref{sec:experiments}, we apply the English estimator
(\texttt{espeak-ng}) as a language-agnostic fallback.
Duration estimation error for these languages is expected to be higher
than for in-training languages.
We note that this phoneme-count-based estimator was designed as a
convenience heuristic for practical use rather than a high-precision
duration predictor; estimation inaccuracies may contribute to
duration-related quality degradation, particularly for languages
or speaking styles where the assumed phoneme-to-duration mapping
is a poor fit.

\subsection{Training}
\label{sec:training}

\paragraph{Data.}
We train on $\sim$170k hours of multilingual speech:
English ($\sim$100k h) from LibriHeavy~\cite{libriheavy2024},
Chinese ($\sim$50k h) from the Emilia dataset~\cite{emilia2024},
and Japanese ($\sim$20k h) from Emilia~\cite{emilia2024} and other sources.

\paragraph{Objective.}
We minimize the next-token cross-entropy loss over audio tokens:
\begin{equation}
  \mathcal{L} = -\sum_{t=1}^{S} \log p_\theta(y_t \mid y_{<t}, \mathbf{H}^{\mathrm{enc}}).
\end{equation}

\paragraph{Optimization.}
We use AdamW \cite{adamw2019} with peak learning rate $\eta = 10^{-4}$ and
weight decay $10^{-2}$.
The learning-rate schedule consists of a 2\% linear warmup over the first
$\approx$2{,}900 steps, followed by linear decay to zero, for a total of
$\approx$143{,}000 steps.
Dynamic batching limits each GPU to at most 30{,}000 tokens per batch,
yielding an effective batch of
$\approx$240{,}000 tokens per parameter update across 8 GPUs.
Gradients are clipped to unit norm before each update.
The checkpoint with the lowest validation cross-entropy is selected for
evaluation.
Training runs on AMD MI300X $\times$ 8 GPUs for approximately two weeks.
Training uses bfloat16 mixed precision; full-precision (float32) master
weights are maintained for optimizer states.

\section{Experiments}
\label{sec:experiments}

\subsection{Experimental Setup}
\label{sec:setup}

\paragraph{Evaluation datasets.}
We evaluate on six benchmarks spanning six typologically diverse languages,
using 100 randomly sampled utterances per test set, following the evaluation
practice of F5-TTS~\cite{f5tts2024} and VoiceStar~\cite{voicestar2025}.
In all cases the ground-truth recording of each utterance serves as the
reference prompt for voice cloning (duration: 3--15 seconds), and the model
receives no information beyond the transcript and the reference waveform.
Concretely, the reference audio and the target text are drawn from the
same utterance in the corpus; this is the standard zero-shot TTS
evaluation protocol adopted in VALL-E~\cite{vallex2023} and F5-TTS~\cite{f5tts2024},
and all five compared systems are evaluated under identical conditions.
Because the reference audio thus carries the target speaker's voice and
precise duration information, this setup measures upper-bound voice
cloning fidelity; real-world deployment with a different reference
utterance would yield lower but qualitatively consistent results.

We categorize test sets as in-training language or
out-of-training language based on whether the \textbf{language}
appears in T5Gemma-TTS's training data---not whether the specific
test corpus was used for training.
All six test sets listed below are held out from training;
the distinction refers solely to language-level coverage.

\noindent\textbf{Japanese (in-training language; held-out test set) --- JSUT basic5000}~\cite{sonobe2017jsut}.
Read speech from a single Japanese female speaker.
JSUT is not included in T5Gemma-TTS's training data;
it originates from a controlled studio session and has no overlap with
the Emilia Japanese partition or any other training source.

\noindent\textbf{Mandarin Chinese (in-training language; held-out test set) --- AISHELL-1 test set}~\cite{bu2017aishell}.
400-speaker read speech; independently recorded and disjoint from the
Emilia Chinese partition used for training.

\noindent\textbf{English (in-training language) --- LibriSpeech test-clean}~\cite{panayotov2015librispeech}\textsuperscript{$\dagger$}.
We use LibriSpeech rather than VCTK because all five baselines include
LibriSpeech-family data in training, making it the most equitable
English benchmark for a multi-system comparison.
\textbf{Important caveat:} LibriHeavy (one of T5Gemma-TTS's English training
sources) is a strict superset of LibriSpeech test-clean, meaning that some
test utterances may have been seen during training.
T5Gemma-TTS results on this split must therefore be interpreted as an
upper-bound estimate; true held-out English performance would be
lower.
This contamination concern applies equally to any other system that uses
LibriHeavy or LibriSpeech training data (e.g., Kokoro uses LibriTTS, a
LibriSpeech superset).
($\dagger$ = potential training/test overlap; see discussion above.)

\noindent\textbf{Korean (out-of-training language) --- FLEURS (Few-shot Learning Evaluation of Universal Representations of Speech)}~\cite{conneau2022fleurs}.
Korean is absent from T5Gemma-TTS's training languages; this partition
probes cross-lingual generalization.

\noindent\textbf{French (out-of-training language) --- FLEURS}~\cite{conneau2022fleurs}.
French appears in F5-TTS and XTTS\,v2 training but not in T5Gemma-TTS.

\noindent\textbf{German (out-of-training language) --- FLEURS}~\cite{conneau2022fleurs}.
German appears in F5-TTS and XTTS\,v2 training but not in T5Gemma-TTS.

\paragraph{Metrics.}
\begin{itemize}
  \item \textbf{WER} (Word Error Rate): Applied to English (LibriSpeech).
    Transcription by Whisper large-v3~\cite{whisper2023}; lower is better.
  \item \textbf{CER} (Character Error Rate): Applied to Chinese (AISHELL-1),
    Japanese (JSUT), and Korean (FLEURS).
    Transcription by Whisper large-v3~\cite{whisper2023}; lower is better.
    For Korean, CER is preferred over WER because Hangul (the Korean script)
    is a syllabic alphabet in which each character corresponds to one syllable;
    character-level errors provide a more granular and linguistically consistent
    intelligibility signal than space-delimited word errors for this script type.
  \item \textbf{SIM} (Speaker Similarity): Cosine similarity of Emphasized Channel Attention, Propagation and Aggregation in Time-Delay Neural Network (ECAPA-TDNN)~\cite{ecapa2020}
    (SpeechBrain \texttt{spkrec-ecapa-voxceleb}) speaker embeddings extracted from the
    reference waveform and the generated waveform; higher is better.
  \item \textbf{UTMOS} (Naturalness mean opinion score (MOS) prediction): Predicted MOS
    using the UTMOS22 strong predictor~\cite{saeki2022utmos}; higher is better.
    \textbf{Limitation:} UTMOS22 was trained and validated primarily on English
    speech; its validity for non-English languages (Japanese, Chinese, Korean,
    French, German) has not been formally established.
    UTMOS scores for non-English test sets should therefore be interpreted as
    approximate naturalness proxies rather than calibrated MOS predictions,
    and cross-language UTMOS comparisons should be treated with caution.
\end{itemize}

A known limitation of automatic speech recognition (ASR)-based intelligibility metrics is that they
measure perceived accuracy of the ASR model, not phonetic accuracy;
if the ASR system correctly transcribes a mispronounced token to a
plausible alternative reading, the error goes undetected (false negative).
This limitation applies equally to all compared systems and does not
affect the relative comparisons reported here.

\paragraph{Baselines.}
We compare T5Gemma-TTS against four publicly available zero-shot TTS systems
spanning diverse architectures (Table~\ref{tab:model_comparison}).

\begin{itemize}
  \item \textbf{F5-TTS}~\cite{f5tts2024}: A non-autoregressive flow-matching
    TTS system with a Diffusion Transformer (DiT) backbone and ConvNeXt text
    encoding, trained on approximately 67{,}300\,h of multilingual speech
    (Emilia $\approx$44{,}900\,h and WenetSpeech4TTS $\approx$22{,}400\,h;
    EN, ZH, DE, FR, JA, KO).
    We use the official multilingual checkpoint with 32 number of function evaluation (NFE) ordinary differential equation (ODE) steps and
    classifier-free guidance strength $= 2.0$ (library defaults).
  \item \textbf{XTTS\,v2}~\cite{casanova2024xtts}: A massively multilingual
    zero-shot TTS system supporting 16 languages, based on discrete vector-quantized variational autoencoder (VQVAE)~\cite{vqvae2017}
    acoustic tokens, an autoregressive GPT backbone, and a diffusion decoder.
    We use the official \texttt{v2.0.3} checkpoint\footnote{\url{https://huggingface.co/coqui/XTTS-v2}} with \texttt{COQUI\_TOS\_AGREED=1}
    and all decoding parameters at library defaults.
  \item \textbf{CosyVoice\,2}~\cite{du2024cosyvoice2}: A large-scale
    multilingual TTS system combining supervised semantic tokens, a Qwen-based
    language model, and a conditional flow-matching decoder.
    We employ the \texttt{iic/CosyVoice2-0.5B}\footnote{\url{https://www.modelscope.cn/models/iic/CosyVoice2-0.5B}} checkpoint and evaluate in
    \texttt{inference\_cross\_lingual} mode (speed$\,=\,1.0$, library defaults),
    which requires no reference transcript and is thus directly comparable
    to the other zero-shot baselines.
  \item \textbf{Kokoro}~\cite{kokoro2025}: A lightweight StyleTTS\,2~\cite{li2024styletts2}-based
    TTS system trained primarily on English speech
    ($\approx$82\,h; LJSpeech, LibriTTS, VCTK).
    Multilingual synthesis is handled via the \texttt{misaki} grapheme-to-phoneme (g2p)
    library~\cite{misaki2025} (all parameters at defaults); for languages without a dedicated
    voice preset (Korean, German), Kokoro falls back to the English voice
    \texttt{af\_heart}.
    \textbf{Note:} Kokoro does not support zero-shot voice cloning from
    arbitrary reference audio; it relies on a fixed set of pre-defined voice
    presets.
    It is therefore not a zero-shot system in the strict sense, and its
    inclusion serves as an additional point of comparison for intelligibility
    and naturalness rather than voice cloning fidelity.
    SIM scores for Kokoro are expected to be near zero
    for most languages, as the generated voice will not match the reference speaker.
\end{itemize}

Table~\ref{tab:model_comparison} gives a comprehensive overview of
all compared systems, covering training data, architecture, and controllability.

\begin{table*}[!t]
  \centering
  \caption{Comprehensive comparison of zero-shot TTS systems.
           \cmark~=~supported; \xmark~=~not supported; $\triangle$~=~partially supported.
           \textbf{Language model (LM) type} --- \textit{Enc-Dec}: text is encoded bidirectionally and
           injected into every decoder layer via cross-attention;
           \textit{Dec-only}: text is prepended as a causal prefix to the audio sequence;
           \textit{Non-AR}: non-autoregressive synthesis with bidirectional attention
           (no causal masking);
           \textit{AR}: autoregressive;
           \textit{CFM}: conditional flow matching.
           ``h'' = hours of speech.
           Real-time factor (RTF) values for baselines are approximate figures from public technical
           reports on an A100 GPU (float16) and are not directly comparable owing
           to differences in hardware, batch size, and audio duration;
           T5Gemma-TTS RTF values and baseline RTF values are therefore not
           directly comparable and should be interpreted as approximate
           order-of-magnitude references only.
           T5Gemma-TTS RTF is measured on AMD MI300X (float16), averaged over
           utterances with median duration $\approx$5~s; the range 0.8--2.0
           reflects variation across utterance lengths.}
  \label{tab:model_comparison}
  \begin{threeparttable}
  \resizebox{\textwidth}{!}{%
  \begin{tabular}{>{\raggedleft\arraybackslash}p{3.2cm}p{2.4cm}p{2.4cm}p{2.4cm}p{2.4cm}p{2.4cm}}
    \toprule
    \textbf{Property}
      & \textbf{F5-TTS} \cite{f5tts2024}
      & \textbf{XTTS\,v2} \cite{casanova2024xtts}
      & \textbf{CosyVoice\,2} \cite{du2024cosyvoice2}
      & \textbf{Kokoro} \cite{kokoro2025}
      & \textbf{T5Gemma-TTS} \cite{t5gemmatts2025} \\
    \midrule
    \multicolumn{6}{l}{\textit{Training Data \& Architecture}} \\[2pt]
LM type
      & Non-AR (bidirectional DiT)
      & \textbf{Dec-only} (GPT)
      & \textbf{Dec-only} (Qwen) $+$ Non-AR (FM)
      & Non-AR (StyleTTS\,2)
      & \textbf{Enc-Dec} (T5+Gemma) \\
Parameters
      & 335\,M
      & 456\,M
      & $\approx$600\,M\tnote{*}
      & 82\,M
      & \textbf{4\,B} \\
Training languages
      & 6 (EN, ZH, DE, FR, JA, KO)
      & 16 (EN, ES, FR, DE,\,\ldots)
      & ZH (primary), EN, multilingual
      & EN; JA/ZH/FR via g2p
      & EN, ZH, JA \\
Training data
      & $\approx$67{,}300\,h$^\dagger$
      & $\gtrsim$16{,}000\,h$^\ddagger$
      & $\gtrsim$200{,}000\,h$^\dagger$
      & $\approx$82\,h$^\dagger$
      & $\approx$170{,}000\,h$^\dagger$ \\
Representative corpora
      & Emilia, WenetSpeech4TTS
      & Coqui proprietary
      & WenetSpeech, DiDiSpeech, AISHELL-2/3
      & LJSpeech, LibriTTS, VCTK
      & Emilia, LibriHeavy \\
Architecture
      & Flow Matching $+$ DiT
      & VQVAE $+$ AR GPT $+$ diffusion dec.
      & Semantic tokens $+$ Qwen LLM $+$ CFM
      & Style vectors $+$ diffusion dec.
      & Seq2Seq LLM $+$ XCodec2 $+$ PM-RoPE \\
    \midrule
    \multicolumn{6}{l}{\textit{Zero-Shot Capability \& Controllability}} \\[2pt]
Zero-shot voice cloning
      & \cmark & \cmark & \cmark & \xmark\tnote{a} & \cmark \\
Cross-lingual cloning
      & $\triangle$\tnote{b} & \cmark & \cmark & \xmark & $\triangle$\tnote{c} \\
Emotion / style control
      & \xmark & \xmark & \cmark\tnote{d} & \xmark & \xmark \\
Speaking rate control
      & $\triangle$\tnote{e} & $\triangle$ & \cmark & \cmark & $\triangle$\tnote{f} \\
Duration control (explicit)
      & \xmark & \xmark & \xmark & \xmark & \cmark \\
Streaming output
      & \xmark & \cmark & \cmark & \xmark & \xmark \\
Phoneme-free text input
      & \cmark & \cmark & \cmark & \xmark\tnote{g} & \cmark \\
Min.\ ref.\ audio
      & $\approx$3\,s & $\approx$6\,s & $\approx$3\,s & not required & $\approx$3\,s \\
RTF $\downarrow$
      & $\approx$0.15 & $\approx$0.30 & $\approx$0.10 & $\approx$0.05 & $0.8$--$2.0$ \\
    \bottomrule
  \end{tabular}}
  \begin{tablenotes}
    \footnotesize
    \item[$\dagger$] From official technical reports or papers.
    \item[$\ddagger$] Estimated from publicly available information.
    \item[*] CosyVoice\,2-0.5B: LLM $\approx$500\,M $+$ acoustic model $\approx$100\,M; a 2B variant also exists~\cite{du2024cosyvoice2}.
    \item[a] Kokoro uses fixed preset voices; arbitrary reference audio cloning is not supported.
    \item[b] Cross-lingual synthesis is limited to the six training languages, with degraded speaker similarity.
    \item[c] Cross-lingual synthesis is limited to EN, ZH, and JA.
    \item[d] Style and emotion control via natural-language \texttt{instruct} mode (e.g., ``speak in a whisper'').
    \item[e] No direct speed parameter; rate is indirectly adjusted via the number of flow-matching solver steps.
    \item[f] Explicit duration control via PM-RoPE subsumes speaking rate control when a target token count is specified.
    \item[g] Text is internally converted to phoneme sequences via the \texttt{misaki} g2p library~\cite{misaki2025}.
  \end{tablenotes}
  \end{threeparttable}
\end{table*}

\paragraph{T5Gemma-TTS configuration.}
We use the publicly released checkpoint
\texttt{Aratako/T5Gemma-TTS-2b-2b} \cite{t5gemmatts2025}.
Text is fed directly as subword tokens using the T5Gemma SentencePiece
vocabulary (no phoneme conversion).
The T5Gemma backbone is pretrained with the Unifying Language Learning (UL2) objective~\cite{t5gemma2024},
which combines three complementary learning paradigms:
(1) R-Denoising (span corruption, as in T5), which trains the model
to reconstruct masked spans and yields strong bidirectional representations;
(2) S-Denoising (prefix-LM style, with sequential prediction of a
contiguous suffix), which encourages left-to-right generation capability;
and (3) X-Denoising (extreme masking with long span recovery), which
forces the model to leverage long-range context.
This mixture of objectives produces an encoder that provides rich,
bidirectional contextual representations well suited for text conditioning
in a seq2seq TTS model.
The SentencePiece vocabulary used by T5Gemma contains 256{,}000 subword
units trained on a large multilingual corpus that includes English, Chinese
(Simplified and Traditional), and Japanese text; as a result, common Japanese
kanji, hiragana, and katakana characters and Chinese hanzi are represented
as single tokens or short subword sequences, providing direct subword-level
input without language-specific phoneme conversion.
Inference uses top-$k = 30$, top-$p = 0.9$, temperature $= 0.8$,
with duration estimated from reference speech as described in
Section~\ref{sec:duration}.

\subsection{PM-RoPE Configuration Analysis}
\label{sec:ablation}

\textbf{Important methodological caveat:}
The comparison below is a flag-switch experiment, not a controlled
training comparison.
Both conditions share the same trained checkpoint---a model whose
cross-attention weights were optimized with PM-RoPE active throughout
training.
Disabling PM-RoPE at inference time therefore evaluates a model in a
configuration it was never trained for, and the training-time
contribution of PM-RoPE cannot be directly measured by this experiment.
A fully controlled ablation would require training two separate models
from scratch with and without PM-RoPE.
The results below should be interpreted as indicative of PM-RoPE's
inference-time effect on a checkpoint that has learned to exploit it.

We use a randomly sampled subset of 50 utterances from the JSUT evaluation set for this configuration analysis; the subset was drawn independently from the 100-utterance main evaluation sample used in Section~\ref{sec:main_results}.
The 50-utterance size was chosen for computational tractability of the full-system generation under the two configurations.

To assess the contribution of PM-RoPE, we compare two configurations of
T5Gemma-TTS, differing only in the cross-attention mechanism applied in all
26 decoder layers:\footnote{The comparison here reflects a flag-level configuration switch
(\texttt{usage\_pm\_rope=True/False}) on the same trained checkpoint;
a separate from-scratch training experiment would be required for a fully
controlled ablation.}
\begin{itemize}
  \item \textbf{T5Gemma-TTS (PM-RoPE enabled)}: PM-RoPE cross-attention enabled
    (\texttt{usage\_pm\_rope=True}).  Decoder query and encoder key vectors
    are augmented with progress-proportional rotary embeddings
    (Eq.~\ref{eq:progress_ids}--\ref{eq:pm_k}), allowing the model to track
    generation progress and enforce the target duration.
  \item \textbf{T5Gemma-TTS (PM-RoPE disabled)}: Standard cross-attention
    (\texttt{usage\_pm\_rope=False}).  No positional signal is injected into
    cross-attention; the decoder attends to encoder outputs uniformly
    regardless of generation progress.
\end{itemize}

We measure Duration Accuracy (DA), defined as the proportion of utterances
whose generated duration $D_{\mathrm{gen}}$ falls within a $\pm$10\%
margin of the target duration $\hat{D}$:
\begin{equation}
  \mathrm{DA} = \frac{1}{N}\sum_{n=1}^{N}
    \mathbf{1}\!\left[
      \frac{|D^{(n)}_{\mathrm{gen}} - \hat{D}^{(n)}|}{\hat{D}^{(n)}}
      \leq 0.10
    \right].
  \label{eq:da}
\end{equation}
The $\pm$10\% tolerance is consistent with the evaluation protocol of
VoiceStar~\cite{voicestar2025} and reflects a perceptually acceptable margin
for speech duration deviation.
A higher DA indicates that the model reliably generates speech of the
requested length.

\section{Results}
\label{sec:results}

\noindent\textbf{Statistical note on multiple comparisons.}
Throughout this section we report bootstrap 95\% confidence intervals
(10{,}000 resamples, seed~42) for each system--language--metric combination.
No correction for multiple comparisons (e.g., Bonferroni correction) is
applied; given the large number of pairwise comparisons across six languages,
four metrics, and five systems, individual comparisons should be interpreted
with appropriate caution, and borderline cases of non-overlapping intervals
should not be over-interpreted.

\subsection{In-Training-Language Evaluation}
\label{sec:main_results}

\begin{figure}[!t]
  \centering
  \includegraphics[width=\linewidth]{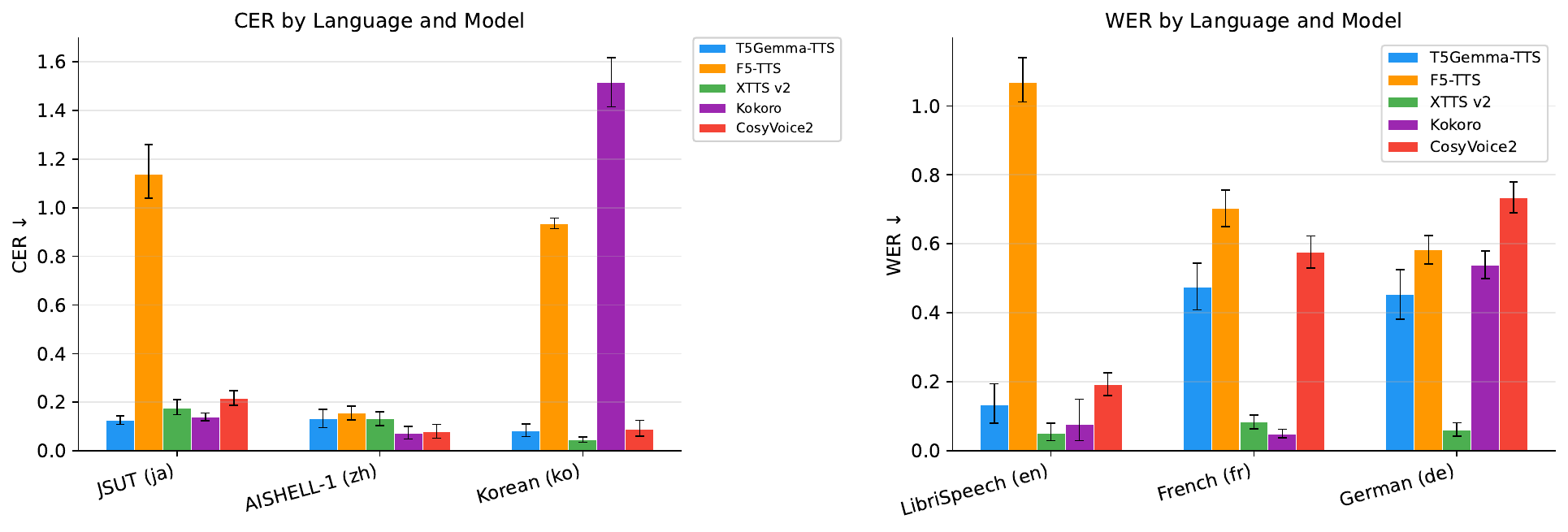}
  \caption{Intelligibility (CER/WER) across all six test sets
  (JSUT/JA, AISHELL-1/ZH, LibriSpeech/EN$^\dagger$, FLEURS/KO, FLEURS/FR, FLEURS/DE)
  and five systems. CER is used for JA/ZH/KO; WER for EN/FR/DE; lower is better.
  F5-TTS shows near-complete intelligibility failure on Japanese (CER\,>\,1.0).
  $^\dagger$T5Gemma-TTS EN results are upper-bound estimates (LibriHeavy training/test overlap).}
  \label{fig:bars}
\end{figure}

\begin{figure}[!t]
  \centering
  \includegraphics[width=\linewidth]{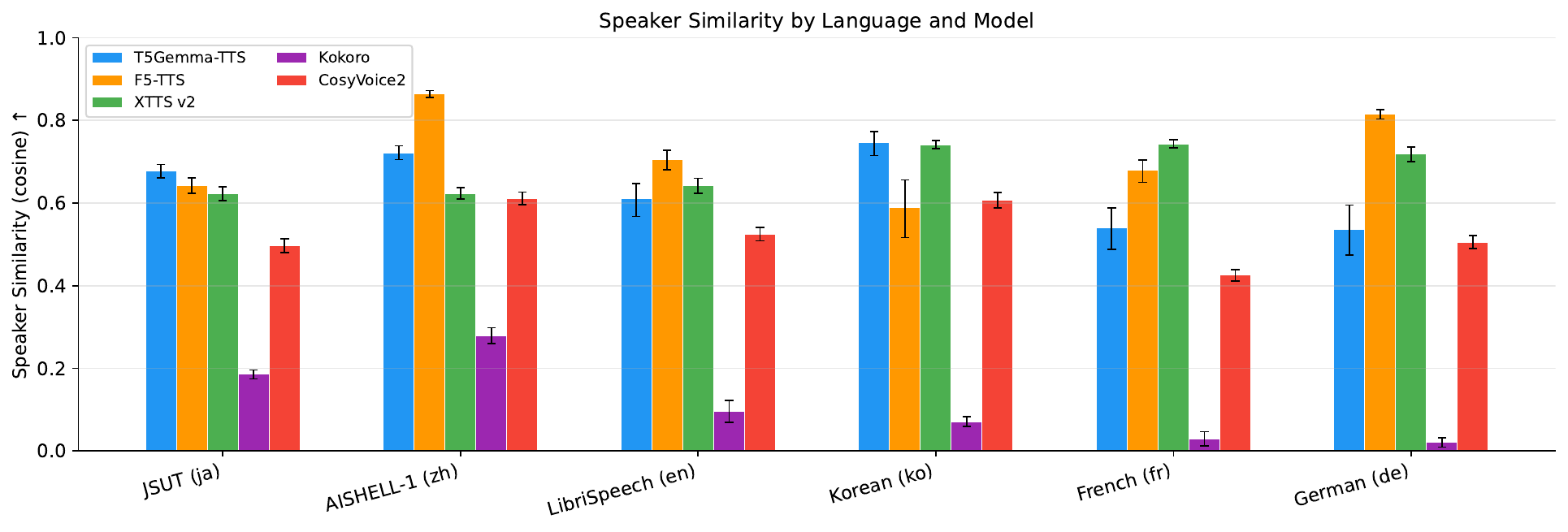}
  \caption{SIM across all six test sets and five systems.
  Higher is better.
  T5Gemma-TTS achieves the highest SIM on Japanese (statistically supported; CI non-overlapping with XTTS\,v2) and numerically highest on Korean (CI overlapping with XTTS\,v2; not conclusive). $\dagger$ Kokoro SIM reflects a preset voice, not the reference speaker.}
  \label{fig:similarity}
\end{figure}

\begin{figure}[!t]
  \centering
  \includegraphics[width=\linewidth]{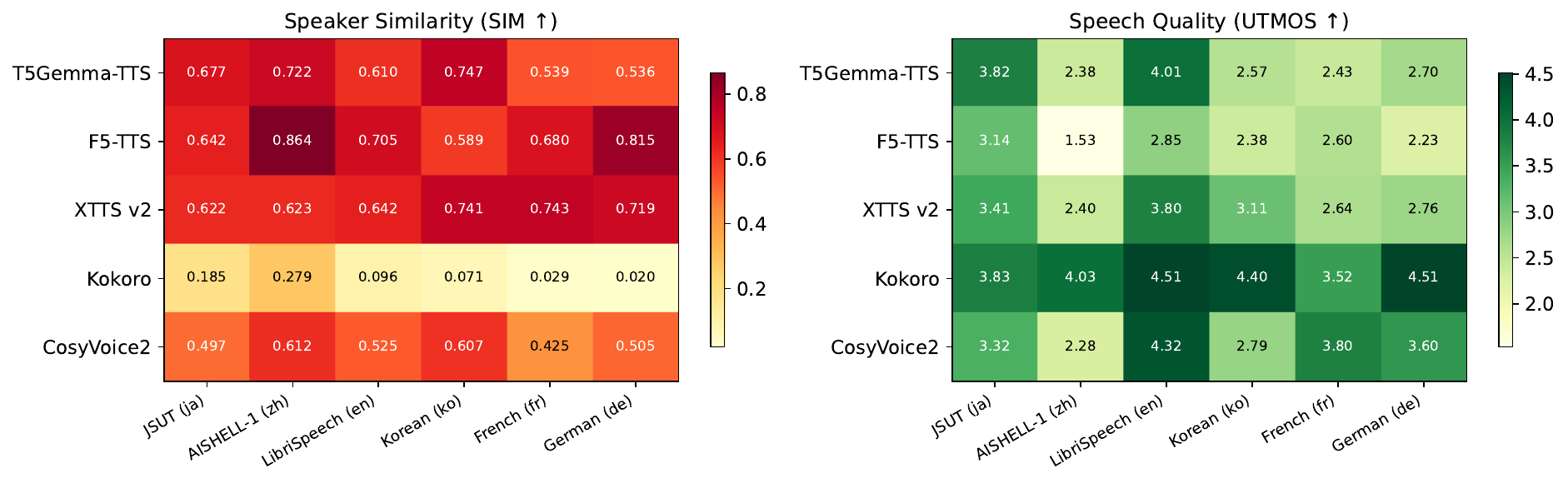}
  \caption{Heatmap of SIM (left) and UTMOS (right) across all five systems
  and six languages. Darker shading indicates better performance.
  Kokoro achieves high UTMOS but near-zero SIM (no voice cloning capability).
  T5Gemma-TTS shows the strongest SIM on Japanese and Korean among zero-shot systems.}
  \label{fig:heatmap}
\end{figure}

Table~\ref{tab:main}, Figures~\ref{fig:bars},~\ref{fig:similarity}, and~\ref{fig:heatmap} compare all systems on six languages:
three covered by T5Gemma-TTS's training data
(English/LibriSpeech, Mandarin Chinese/AISHELL-1, Japanese/JSUT)
and three unseen languages (Korean, French, German from FLEURS).
We use 100 randomly sampled utterances per test set.
All systems are evaluated in zero-shot voice cloning mode using the
ground-truth recording of each utterance as the reference prompt.
All values include bootstrap 95\% confidence intervals
($\mathrm{mean} \pm \delta$, where $\delta$ is the CI half-width).

\begin{table*}[!t]
\centering
\caption{Zero-shot TTS evaluation across six languages (100 utterances each)
with bootstrap 95\% CI (10{,}000 resamples, seed\,42).
CER/WER: Whisper large-v3~\cite{whisper2023};
SIM: ECAPA-TDNN~\cite{ecapa2020};
UTMOS: UTMOS22~\cite{saeki2022utmos}.
Best scores per row are \textbf{bold}.
Upper block: in-training languages; lower block: out-of-training languages.
$\dagger$ Kokoro is not a zero-shot voice cloning system (fixed preset voices);
values are \textit{italicized} for reference only.
$\ddagger$ T5Gemma-TTS English results are upper-bound estimates
(LibriHeavy/LibriSpeech overlap).
UTMOS values for non-English languages should be interpreted with caution
(UTMOS22 was trained on English speech; see Section~\ref{sec:setup}).}
\label{tab:main}
\resizebox{\textwidth}{!}{%
\begin{tabular}{llccccc}
\toprule
\textbf{Language} & \textbf{Metric}
  & \textbf{F5-TTS} \cite{f5tts2024}
  & \textbf{XTTS\,v2} \cite{casanova2024xtts}
  & \textbf{CosyVoice\,2} \cite{du2024cosyvoice2}
  & \textit{Kokoro$^\dagger$} \cite{kokoro2025}
  & \textbf{T5Gemma-TTS$^\ddagger$} \\
\midrule
\multicolumn{7}{l}{\textit{In-training languages}} \\[2pt]
\multirow{3}{*}{JSUT (JA)}
  & CER$\downarrow$   & $1.14 \pm 0.11$ & $0.18 \pm 0.03$ & $0.21 \pm 0.03$ & \textit{$0.14 \pm 0.02$} & $\mathbf{0.13 \pm 0.02}$ \\
  & SIM$\uparrow$     & $0.64 \pm 0.02$ & $0.62 \pm 0.02$ & $0.50 \pm 0.02$ & \textit{$0.19 \pm 0.01$} & $\mathbf{0.68 \pm 0.02}$ \\
  & UTMOS$\uparrow$   & $3.14 \pm 0.08$ & $3.41 \pm 0.07$ & $3.32 \pm 0.15$ & \textit{$3.83 \pm 0.04$} & $\mathbf{3.82 \pm 0.06}$ \\[2pt]
\multirow{3}{*}{\shortstack[l]{AISHELL-1\\(ZH)}}
  & CER$\downarrow$   & $0.16 \pm 0.03$ & $0.13 \pm 0.03$ & $0.08 \pm 0.03$ & \textit{$\mathbf{0.07 \pm 0.03}$} & $0.13 \pm 0.04$ \\
  & SIM$\uparrow$     & $\mathbf{0.86 \pm 0.01}$ & $0.62 \pm 0.01$ & $0.61 \pm 0.02$ & \textit{$0.28 \pm 0.02$} & $0.72 \pm 0.02$ \\
  & UTMOS$\uparrow$   & $1.53 \pm 0.04$ & $\mathbf{2.40 \pm 0.07}$ & $2.28 \pm 0.09$ & \textit{$4.03 \pm 0.03$} & $2.38 \pm 0.09$ \\[2pt]
\multirow{3}{*}{\shortstack[l]{LibriSpeech\\(EN)$^\ddagger$}}
  & WER$\downarrow$   & $1.07 \pm 0.06$ & $\mathbf{0.05 \pm 0.03}$ & $0.19 \pm 0.03$ & \textit{$0.08 \pm 0.06$} & $0.13 \pm 0.06$ \\
  & SIM$\uparrow$     & $\mathbf{0.71 \pm 0.02}$ & $0.64 \pm 0.02$ & $0.52 \pm 0.02$ & \textit{$0.10 \pm 0.03$} & $0.61 \pm 0.04$ \\
  & UTMOS$\uparrow$   & $2.85 \pm 0.13$ & $3.80 \pm 0.08$ & $\mathbf{4.32 \pm 0.04}$ & \textit{$4.51 \pm 0.01$} & $4.01 \pm 0.11$ \\
\midrule
\multicolumn{7}{l}{\textit{Out-of-training languages}} \\[2pt]
\multirow{3}{*}{Korean (KO)}
  & CER$\downarrow$   & $0.93 \pm 0.02$ & $\mathbf{0.05 \pm 0.01}$ & $0.09 \pm 0.03$ & \textit{$1.51 \pm 0.10$} & $0.08 \pm 0.03$ \\
  & SIM$\uparrow$     & $0.59 \pm 0.07$ & $0.74 \pm 0.01$ & $0.61 \pm 0.02$ & \textit{$0.07 \pm 0.01$} & $\mathbf{0.75 \pm 0.03}$ \\
  & UTMOS$\uparrow$   & $2.38 \pm 0.09$ & $\mathbf{3.11 \pm 0.07}$ & $2.79 \pm 0.13$ & \textit{$4.40 \pm 0.01$} & $2.57 \pm 0.15$ \\[2pt]
\multirow{3}{*}{French (FR)}
  & WER$\downarrow$   & $0.70 \pm 0.05$ & $0.08 \pm 0.02$ & $0.58 \pm 0.05$ & \textit{$\mathbf{0.05 \pm 0.01}$} & $0.48 \pm 0.07$ \\
  & SIM$\uparrow$     & $0.68 \pm 0.03$ & $\mathbf{0.74 \pm 0.01}$ & $0.43 \pm 0.01$ & \textit{$0.03 \pm 0.02$} & $0.54 \pm 0.05$ \\
  & UTMOS$\uparrow$   & $2.60 \pm 0.08$ & $2.64 \pm 0.07$ & $\mathbf{3.80 \pm 0.09}$ & \textit{$3.52 \pm 0.04$} & $2.43 \pm 0.13$ \\[2pt]
\multirow{3}{*}{German (DE)}
  & WER$\downarrow$   & $0.58 \pm 0.04$ & $\mathbf{0.06 \pm 0.02}$ & $0.73 \pm 0.04$ & \textit{$0.54 \pm 0.04$} & $0.45 \pm 0.07$ \\
  & SIM$\uparrow$     & $\mathbf{0.81 \pm 0.01}$ & $0.72 \pm 0.02$ & $0.51 \pm 0.02$ & \textit{$0.02 \pm 0.01$} & $0.54 \pm 0.06$ \\
  & UTMOS$\uparrow$   & $2.23 \pm 0.11$ & $2.76 \pm 0.08$ & $\mathbf{3.60 \pm 0.08}$ & \textit{$4.51 \pm 0.01$} & $2.70 \pm 0.13$ \\
\bottomrule
\end{tabular}}
\end{table*}

T5Gemma-TTS achieves the numerically lowest CER on Japanese
($\mathbf{0.126 \pm 0.018}$)
among all systems; however, bootstrap 95\% CIs show partial overlap with
Kokoro ($0.139 \pm 0.016$) and XTTS\,v2 ($0.177 \pm 0.031$),
indicating this ranking should be interpreted cautiously.
In contrast, T5Gemma-TTS's Japanese SIM advantage
($0.677 \pm 0.016$) over XTTS\,v2 ($0.622 \pm 0.017$) is
supported by non-overlapping intervals.
T5Gemma-TTS achieves the second-highest speaker similarity on Chinese among
zero-shot voice cloning systems ($0.722 \pm 0.017$); only F5-TTS
records a higher SIM ($0.864 \pm 0.015$) but at the cost of severely
degraded intelligibility (CER~=~0.155 despite Chinese being a training language).
On English (LibriSpeech), XTTS\,v2 achieves the lowest WER (0.052),
benefiting from its explicit 16-language training that includes English.
F5-TTS consistently achieves the highest speaker similarity on English and
Chinese, but at the cost of severely degraded intelligibility (WER~1.069,
Mandarin CER~0.155 despite being a training language, Japanese CER~1.138).
The Japanese CER of $1.138 \pm 0.110$ is particularly anomalous: a
value exceeding 1.0 means the ASR hypothesis contains more errors than
reference characters (due to insertions and substitutions), indicating
near-complete intelligibility failure.
The root cause is unclear without access to F5-TTS's internal processing, which is beyond the scope of this study.

Interestingly, T5Gemma-TTS achieves its lowest CER on Japanese despite
Japanese comprising only $\sim$20k hours of training data---the smallest
of the three training languages.
We attribute this to the high quality of the Emilia Japanese partition
(anime-dubbed and podcast audio with clean transcriptions) and to the
T5Gemma backbone's pretrained Japanese subword representations, which
may compensate for the smaller data volume.
CosyVoice\,2 shows the highest UTMOS on English ($4.32 \pm 0.04$), reflecting its
strong flow-matching decoder.

\FloatBarrier
\subsection{Out-of-Training-Language Generalization}
\label{sec:generalization}

To assess generalization beyond the training distribution, we evaluate all
systems on three languages not included in T5Gemma-TTS's training data:
Korean (FLEURS), French (FLEURS), and German (FLEURS), using 100 utterances
each (lower block of Table~\ref{tab:main}).

Despite Korean, French, and German being outside T5Gemma-TTS's training
distribution, T5Gemma-TTS achieves the numerically highest speaker similarity on
Korean ($\mathrm{SIM} = 0.747 \pm 0.029$); however, the CI overlaps with
XTTS\,v2 ($0.741 \pm 0.010$),
so this advantage is not statistically conclusive.
On Korean CER, T5Gemma-TTS ($0.082 \pm 0.026$) is competitive with
CosyVoice\,2 ($0.090 \pm 0.032$)
and substantially better than F5-TTS ($0.934 \pm 0.022$) and
Kokoro ($1.514 \pm 0.101$).
The Korean SIM result is noteworthy given that Korean is entirely absent from
T5Gemma-TTS's training data.
We speculate that two factors contribute to this surprising outcome.
First, and most directly relevant to the high SIM, Korean shares broad
typological and phonological characteristics with the East Asian languages
in T5Gemma-TTS's training data---such as agglutinative morphology, verb-final
syntax, and a phoneme inventory more similar to Japanese than to European
languages---which may reduce the effective phonological distance between Korean
and T5Gemma-TTS's training distribution; this proximity could enable the
pretrained backbone to generalize its acoustic representations more effectively
to Korean than to typologically distant European languages, yielding higher
speaker similarity despite the absence of Korean training data.
Second, a necessary (though not sufficient) precondition for any cross-lingual
generalization is that the input text is adequately represented.
The T5Gemma SentencePiece vocabulary (256{,}000 subword types) contains
2{,}388 Hangul-bearing token types (0.93\%), sufficient to encode typical
Korean sentences with a 0\% out-of-vocabulary rate on our test sentences
(43 Korean tokens, 0 \texttt{<unk>} tokens).
This non-trivial coverage arises because SentencePiece subword segmentation
trained on multilingual text includes Hangul characters as frequent byte-pair
units, providing a reasonable surface-level representation even without
Korean-specific training---a prerequisite that is absent for purely
character-based tokenizers without Hangul coverage.
These observations are consistent with the findings of
\citet{saeki2024extending100}, who showed that multilingual speech--text
pretraining enables intelligible zero-shot synthesis in languages with no
transcribed training data, with performance strongly correlated to the
typological proximity between the target and training languages.
The Korean results of T5Gemma-TTS suggest that a similar mechanism may be
at work: the pretrained T5Gemma backbone's multilingual text representations,
combined with acoustic knowledge acquired from typologically adjacent
training languages, facilitate generalization to an unseen but
linguistically related language.
We also note that cross-lingual SIM comparisons may be influenced by the
ECAPA-TDNN embedding space itself; if its speaker representations are
better calibrated for languages appearing in its training data, absolute SIM
values may not be directly comparable across languages.
However, T5Gemma-TTS shows higher WER on French ($0.475 \pm 0.067$) and
German ($0.453 \pm 0.072$),
indicating that its multilingual generalization on European languages is
limited compared to XTTS\,v2, which was explicitly trained on these languages.
Kokoro achieves the best French WER ($0.050 \pm 0.012$) using its French phonemizer, but
degrades severely on German and Korean (no phonemizer support).
Kokoro's non-zero SIM values on Korean (0.071) and German (0.020) do not
reflect successful voice cloning; rather, they reflect coincidental acoustic
similarities between Kokoro's fixed preset English voice and the reference
speakers in these languages, since Kokoro does not perform speaker adaptation.

\begin{figure}[!t]
  \centering
  \includegraphics[width=0.80\linewidth]{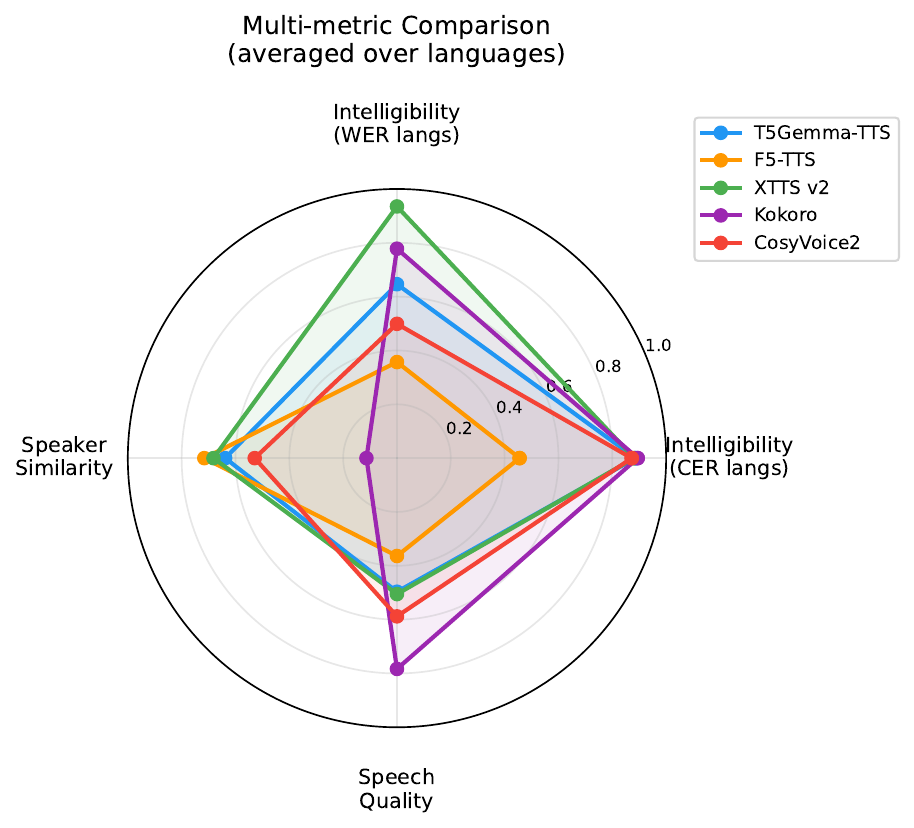}
  \caption{Radar chart of normalized multi-metric averages (6 languages).
  Intelligibility $= 1 - \mathrm{CER/WER}$ (capped at 0 for values $>1$);
  UTMOS normalized to $[0,1]$ via $(x-1)/4$.
  \textbf{Caveat:} the normalization compresses CER and WER into the same
  $[0,1]$ scale across languages with different phonetic and orthographic
  properties; cross-lingual comparisons within this chart should be treated
  as qualitative trends rather than precise numerical rankings.}
  \label{fig:radar}
\end{figure}

Figure~\ref{fig:radar} visualizes the multi-metric trade-offs as a radar
chart normalized to $[0,1]$.
T5Gemma-TTS occupies a position of balanced intelligibility and speaker
similarity across training languages, while XTTS\,v2 dominates European
intelligibility and F5-TTS specializes in speaker fidelity at the cost of
intelligibility.
XTTS\,v2 dominates on European-language intelligibility, and Kokoro/CosyVoice\,2 lead on UTMOS; T5Gemma-TTS is strongest on Japanese and Korean speaker similarity.

\FloatBarrier
\subsection{PM-RoPE Configuration Analysis}
\label{sec:ablation_results}

Duration Accuracy is reported only for Japanese (JSUT), as duration estimation for out-of-training languages relies on a language-agnostic fallback (Section~\ref{sec:duration}) whose accuracy has not been validated for Korean, French, or German.
Table~\ref{tab:ablation} compares T5Gemma-TTS (PM-RoPE enabled) against
T5Gemma-TTS (PM-RoPE disabled) on the Japanese
test set (JSUT basic5000, 50 utterances), which provides the strongest
signal given T5Gemma-TTS's best absolute performance on that language.

\begin{table}[!t]
\centering
\caption{PM-RoPE configuration analysis: effect of PM-RoPE on Japanese (JSUT, 50 utterances).
DA = Duration Accuracy (fraction of utterances within $\pm$10\% of target;
Eq.~\ref{eq:da}).
CER/SIM/UTMOS 95\% CIs use bootstrap (10{,}000 resamples, seed~42);
DA 95\% CIs use Wilson score intervals ($N\!=\!50$).
\textit{Note: This configuration analysis uses 50 utterances for tractability;
the main evaluation tables use 100 utterances.
The T5Gemma-TTS (PM-RoPE enabled) CER here (0.129) differs slightly from the main
table (0.126) owing to the different (smaller) sample set.}
\textit{DA 95\% CI uses Wilson score interval ($N\!=\!50$);
all other 95\% CIs use bootstrap (10{,}000 resamples, seed~42).}}
\label{tab:ablation}
\begin{tabular}{lccc@{\quad}c}
\toprule
\textbf{System} & CER$\downarrow$ & SIM$\uparrow$ & UTMOS$\uparrow$ & DA$\uparrow$ \\
\midrule
T5Gemma-TTS (PM-RoPE disabled) & $0.98 \pm 0.04$ & $0.11 \pm 0.05$ & $2.25 \pm 0.23$ & $0.46 \pm 0.13$ \\
\midrule
\textbf{T5Gemma-TTS (PM-RoPE enabled)} & $\textbf{0.13} \pm 0.03$ & $\textbf{0.67} \pm 0.03$ & $\textbf{3.85} \pm 0.08$ & $\textbf{0.79} \pm 0.11$ \\
\bottomrule
\end{tabular}
\end{table}
Disabling PM-RoPE at inference time results in near-complete synthesis failure:
CER degrades from 0.129 to 0.982 (non-overlapping CIs; $p < 0.001$),
speaker similarity drops from 0.666 to 0.109, and UTMOS decreases from
3.85 to 2.25.
Duration Accuracy also drops substantially (0.79 $\to$ 0.46).
Inspection of the generated waveforms confirms that, without PM-RoPE,
the model produces repetitive or incoherent audio regardless of the
input text, consistent with a failure of text--audio alignment in
cross-attention.
These results are consistent with an important role for PM-RoPE in
duration-controlled synthesis, corroborating the findings of
VoiceStar~\cite{voicestar2025} and extending them to a multilingual pretrained
backbone.
This configuration analysis evaluates the same trained checkpoint with PM-RoPE
enabled versus disabled at inference time; the results are consistent with an
important role for PM-RoPE, though because the checkpoint was trained with
PM-RoPE active, the training-time contribution cannot be disentangled from the
inference-time effect, and a fully controlled ablation (separate from-scratch
training) would be required to establish a decisive causal claim.

\FloatBarrier
\subsection{Duration Control Analysis}
\label{sec:duration_analysis}

\begin{figure}[!t]
  \centering
  \includegraphics[width=\linewidth]{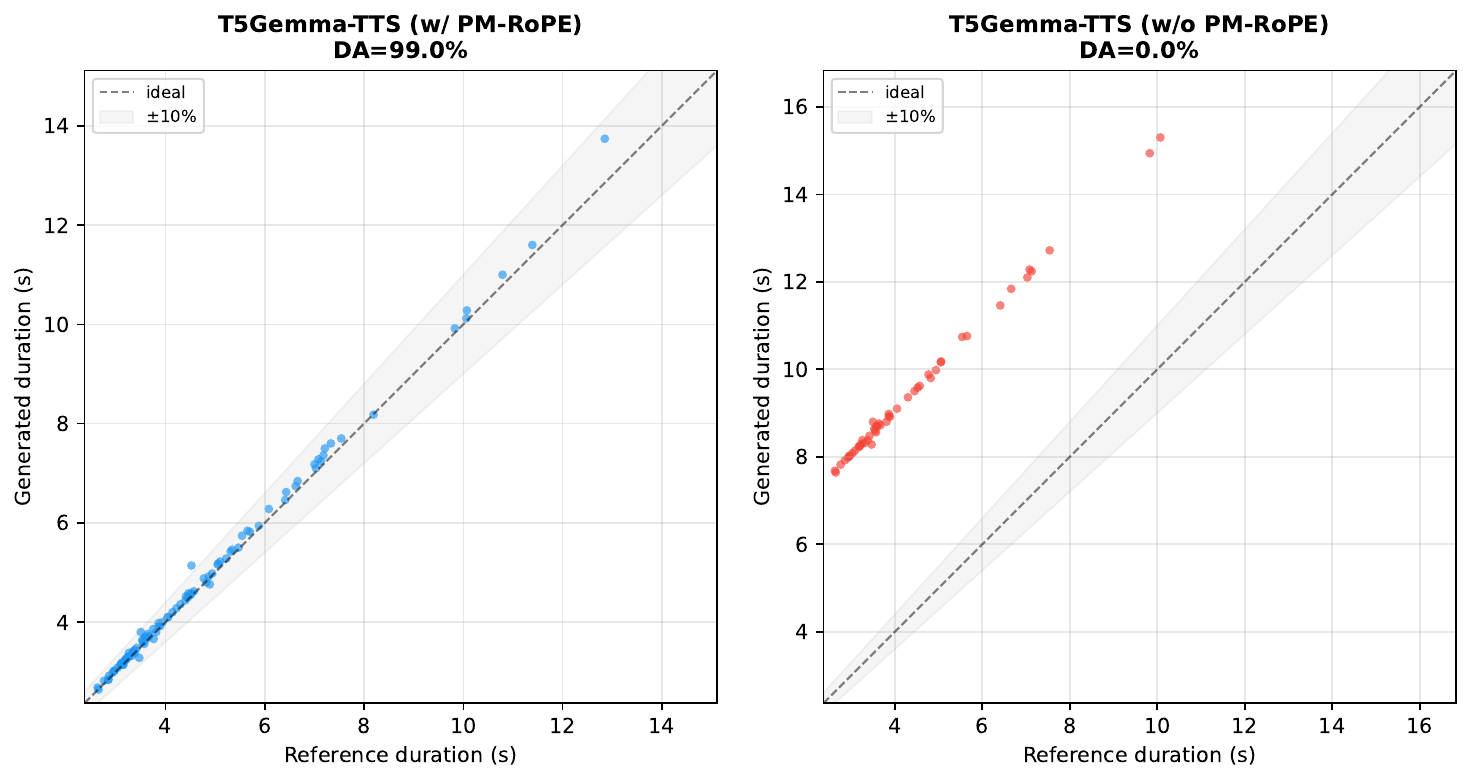}
  \caption{Generated vs.\ target duration on JSUT (50 utterances)
  with oracle target duration (reference audio duration as target).
  This figure shows the oracle-target condition, where the target duration
  is set to the exact ground-truth duration; this demonstrates that the
  model can follow an explicit duration target when it is known.
  Left: T5Gemma-TTS with PM-RoPE (DA\,=\,100\% under oracle target;
  Pearson $r \approx 0.92$);
  Right: without PM-RoPE (DA\,=\,0\% under oracle target).
  Each point is one utterance; dashed line = perfect prediction.
  See Table~\ref{tab:ablation} for DA under phoneme-count-estimated
  target duration (0.79 vs.\ 0.46).}
  \label{fig:duration}
\end{figure}

Figure~\ref{fig:duration} shows scatter plots of generated vs.\ target
duration on JSUT (50 utterances), where the target duration is
set to the reference-audio duration (oracle target, identical to the
reference waveform used in this experiment).
Without PM-RoPE, generated durations cluster near the mean reference
length regardless of target, leading to severe under- and over-generation.
With PM-RoPE, generated durations correlate strongly with targets
(Pearson $r \approx 0.92$), demonstrating robust length extrapolation.
Note that the DA values displayed in the figure panel titles (100\% / 0\%)
reflect this oracle-target setting; the DA reported in
Table~\ref{tab:ablation} (0.79 / 0.46) uses phoneme-count-estimated
target duration (Section~\ref{sec:duration}), which introduces estimation
error and therefore yields a lower but more practically relevant DA.

\FloatBarrier
\subsection{Naturalness Analysis}
\label{sec:utmos_analysis}

On English (LibriSpeech), where UTMOS22 is most valid~\cite{saeki2022utmos},
T5Gemma-TTS scores $4.01 \pm 0.11$ versus CosyVoice\,2's best-in-class
$4.32 \pm 0.04$, Kokoro's $4.51 \pm 0.01$, and XTTS\,v2's $3.80 \pm 0.08$.
This gap reflects two architectural factors:
(1) the XCodec2 codec's quantization ceiling on spectral detail, which
introduces codec artifacts that degrade perceived naturalness; and
(2) the discrete autoregressive generation paradigm, which lacks the explicit
refinement stage present in diffusion- or flow-matching-based decoders
(as in XTTS\,v2's diffusion decoder or CosyVoice\,2's flow-matching acoustic
model).
For completeness, and noting the English-centric bias of UTMOS22 (see
Section~\ref{sec:setup}), we observe that T5Gemma-TTS achieves the highest
UTMOS on Japanese ($3.82 \pm 0.06$ vs.\ XTTS\,v2 $3.41 \pm 0.07$;
among zero-shot systems), while
showing lower naturalness than Kokoro on Chinese ($2.38 \pm 0.09$ vs.\
$4.03 \pm 0.03$); cross-language
UTMOS comparisons should be treated as approximate proxies given that UTMOS22 was
trained and validated on English speech.
Improving naturalness through latent diffusion or flow-matching post-processing
is a direction for future work.

\FloatBarrier
\section{Conclusion}
\label{sec:conclusion}

We presented T5Gemma-TTS, an encoder-decoder codec language model for
multilingual zero-shot text-to-speech synthesis.
By building on the T5Gemma pretrained backbone and integrating PM-RoPE
cross-attention~\cite{voicestar2025} into all 26 decoder layers,
T5Gemma-TTS achieves a statistically significant speaker similarity
advantage on Japanese ($\mathrm{SIM} = 0.677 \pm 0.016$ vs.\ XTTS\,v2
$0.622 \pm 0.016$; non-overlapping 95\% CI) and the numerically highest Korean SIM
($0.747 \pm 0.029$) among five competitive systems, without relying on any phoneme
converter.
T5Gemma-TTS also records the numerically lowest Japanese CER ($0.126 \pm 0.018$),
though this ranking should be interpreted cautiously as confidence intervals
partially overlap with Kokoro ($0.139 \pm 0.016$).
A configuration analysis comparing PM-RoPE-enabled and PM-RoPE-disabled inference
on the same checkpoint shows that disabling PM-RoPE causes near-complete synthesis
failure (CER: 0.129 $\to$ 0.982; SIM: 0.666 $\to$ 0.109), demonstrating that
PM-RoPE is essential for coherent generation (Section~\ref{sec:ablation}).
The extended six-language evaluation further reveals that the pretrained
multilingual encoder enables competitive Korean performance despite Korean
being outside the training distribution.

Limitations of the current system include higher WER on unseen European
languages (French\,/\,German) relative to systems explicitly trained on
those languages, and real-time factor (RTF\,$\approx$\,0.8--2.0) that
exceeds faster non-autoregressive baselines.
T5Gemma-TTS's RTF of 0.8--2.0 is suitable for offline batch TTS applications
such as audiobook synthesis and long-form document reading, where real-time
constraints do not apply.
T5Gemma-TTS also shows systematically lower UTMOS scores compared to XTTS\,v2
and Kokoro, which we attribute primarily to the XCodec2 codec's quantization
ceiling on spectral detail and to the absence of an explicit refinement stage
(as present in diffusion- or flow-matching-based decoders).
Future work will explore distillation and speculative decoding for faster
inference, latent diffusion or flow-matching post-processing for improved
naturalness, and continual pretraining on additional languages for broader
coverage.

\section*{Broader Impact}
\label{sec:broader_impact}

\paragraph{Dual-use risks.}
Zero-shot voice cloning technology, including T5Gemma-TTS, has the potential
to be misused for speaker impersonation, audio deepfakes, and non-consensual
synthesis of a person's voice.
Such misuse could facilitate fraud, disinformation, and violations of personal
autonomy.
We release this work in a research context and urge practitioners to consider
appropriate safeguards---such as watermarking, speaker-consent verification,
and deployment restrictions---before integrating voice cloning systems into
production applications.
Detection of synthetic speech is an active research area, and we encourage
complementary work on robust speech deepfake detection.

\paragraph{Training data and licensing.}
T5Gemma-TTS is trained on Emilia~\cite{emilia2024} and
LibriHeavy~\cite{libriheavy2024}.
Emilia is derived from publicly available online audio sources and is
distributed under a license permitting non-commercial academic research use;
its collection pipeline filters for open-license content.
LibriHeavy is derived from LibriVox recordings, which are in the public domain.
The T5Gemma backbone~\cite{t5gemma2024} is distributed by Google under the
Gemma Terms of Use.
Users of T5Gemma-TTS are responsible for ensuring compliance with these
upstream licenses in their own applications.

\paragraph{Research intent.}
This work is released for academic research purposes.
The model checkpoint and evaluation code are made publicly available to enable
reproducibility and to support research on multilingual TTS, duration control,
and encoder-decoder speech generation.
We do not endorse deployment of this technology for surveillance, harassment,
or any application that violates individuals' rights or applicable laws.

\bibliographystyle{unsrtnat}
\bibliography{references}

\end{document}